\documentclass[12pt]{iopart}

\usepackage{iopams}
\usepackage{graphicx}

\begin{document}

\title{
A Random Laser as a Dynamical Network
}
\author{M. H\"ofner, H.-J. W\"unsche and F. Henneberger}
\address{Humboldt-Universit\"at zu Berlin, Institut f\"ur Physik, Newtonstr. 15, D-12489 Berlin}
\ead{mhoefner@physik.hu-berlin.de}

\begin{abstract} 
The mode dynamics of a random laser is investigated in experiment and theory. The laser consists of a  ZnCdO/ZnO multiple quantum well with air-holes that provide the necessary feedback. Time-resolved measurements reveal multimode spectra with individually developing features but no variation from shot to shot. These findings are qualitatively reproduced with a model that exploits the specifics of a dilute system of weak scatterers and can be interpreted in terms of a lasing network. Introducing the phase-sensitive node coherence reveals new aspects of the self-organization of the laser field. Lasing is carried by connected links between a subset of scatterers, the fields on which are oscillating coherently in phase. In addition, perturbing feedback with possibly unfitting phases from frustrated other scatterers is suppressed by destructive superposition. We believe that our findings are representative at least for weakly scattering random lasers. A generalization to random laser with dense and strong scatterers seems to be possible when using a more complex scattering theory for this case. 
\end{abstract}

%\maketitle

% Intro
\section{Introduction}

Looking at a matter from a different point of view may lead to new insights. Here, we look at a random laser from the point of view of networks. Both subjects are well developed fields of research but have not been brought together so far. The random laser (RL) operates without mirrors or another type of resonator. The necessary optical feedback is provided by multiple scattering of light at inhomogeneities, randomly distributed within the laser medium \cite{caoreview2005,wiersma08,zaitsev10}. Such systems are relatively easy to manufacture and they have interesting applications, among them structure detection of disordered media  \cite{polsonrandom2004,knitterspectro-polarimetric2013,follitime-resolved2013} and speckle-free imaging \cite{reddingspeckle-free2012}. Very different objects can be named a network. A fishnet, a cobweb, the road network, and the internet are well-known examples. What they all have in common is a structure consisting of nodes connected by links. This general concept plays an important role in various branches of science and engineering including subjects as the human brain and even social networks. In optics, networks of multiple coupled lasers are an important example. They represent a specific realization of the generic class of coupled self-sustaining oscillators, offering access to a rich world of dynamical scenarios \cite{Heiligenthal2013,nixon2013}.

A single RL can be deemed as a network in the following sense. Light is alternately scattered at the inhomogeneities and propagated between them. From this point of view, the scatterers are the nodes of a network and the optical pathways between them are the links. With increasing amplification due to stimulated emission, the RL starts to lase when the scattering losses along certain closed pathways of light become compensated by the amplification. The photons on those pathways close to gain-loss cancellation live extremely long, whereas all others decay. They form the so-called lasing modes. This way, lasing of an RL can be regarded as a specific type of self-organizing network. We call it a lasing network because the individual nodes cannot lase but only the network as a whole. It must not be confused with the laser networks mentioned above, where the nodes are individually running lasers coupled by passive links. 

Many RL exhibit co-lasing of multiple modes at seemingly random wavelengths. This feature has been observed in completely different types of RL: a powder of amplifying ZnO particles \cite{caorandom1999}, human tissues \cite{polsonrandom2004}, a passive porous glass filled with a laser dye \cite{mujumdar2007}, a semiconductor chip with scattering air holes \cite{kalus11}, or in a cold-atom RL \cite{Baudouin2013}, to name only few. 

References \cite{Tureci2008a,tureciab2009} give an in-depth theoretical explanation of the multi-mode operation, which is, however, limited to the stationary state. In contrast, practically all experiments are performed under pulsed excitation and mostly time-integrated data are recorded. Under these conditions, it is not clear, whether the different modes indeed coexist at the same time or appear consecutively. 

Taking up these questions, the dynamics of the lasing modes of a RL and the relation to its network structure are the central subjects of the present paper. Experimentally, the lasing modes are identified by  peaks in the optical spectra. Section \ref{sec.experiment} presents time-resolved spectra for a sample similar to that of reference  \cite{kalus11}. Multiple modes coexist at the same times but their relative intensities vary during the excitation pulse. A dynamical model of the RL is presented in section \ref{sec.model} that is able to reproduce relevant qualitative features of the experiment (section \ref{sec.simulation}). The simulation data are used in section \ref{sec.network} to evaluate the RL as a weighted network. In particular, the optical phase is incorporated which reveals that in different modes different parts of the net are excluded from lasing by destructive interference. Finally, the paper is summarized in section \ref{sec.conclusion}.

% Experiment
\section{Experiment: Dynamics of Lasing Modes} \label{sec.experiment}
%---------------------------------------------

\begin{figure}[ht]
\centering
\includegraphics[width=10cm]{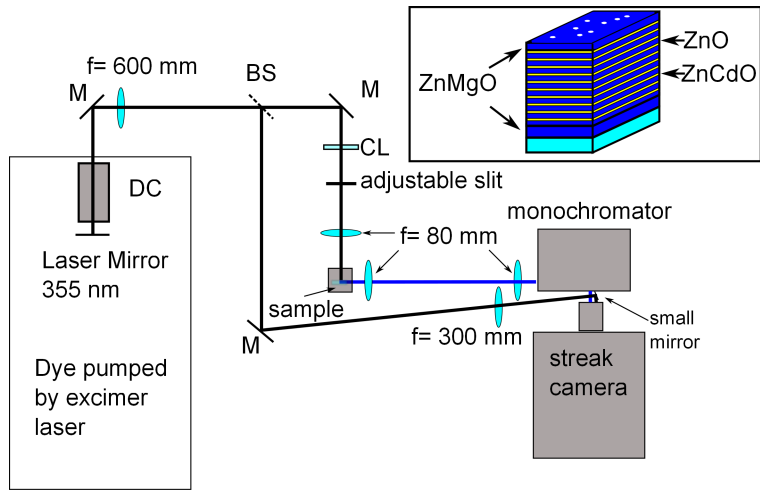}
\caption{ \label{fig:setup}
Experimental setup for time resolved measurement of the random lasing structure based on a 10 ns ultraviolet excitation source (363 nm). M = mirror, BS = beam splitter, f = lens with focal length, CL = Combination of cylindrical and spherical lens, 
DC = Dye cuvette. Panel: Sketch of the sample (no true scale) with the multiple quantum well structure providing the gain and air-holes providing the feedback (buffer layer composition Zn$_{65}$Mg$_{35}$O, quantum well structure Zn$_{88}$Cd$_{12}$O / ZnO). 
}
\end{figure}

%\subsubsection{Sample}
The design of the sample used in the experiment (top panel in figure \ref{fig:setup}) is similar to that sketched in figure 1 of reference  \cite{kalus11}. It is grown on  a-plane sapphire (11-20) beginning with a 650 nm Zn$_\textrm{65}$Mg$_{35}$O buffer layer. The multiple quantum well structure providing the optical gain required for laser action is deposited on this buffer. It consists of ten periods Zn$_{88}$Cd$_{12}$O of 2.6 nm thickness and ZnO of \mbox{7.3 nm} thickness. A cap of \mbox{225 nm} Zn$_{65}$Mg$_{35}$O is grown on top. This layer structure produces a planar waveguide structure, wave propagation can be regarded thus as two-dimensional. Under special conditions in the growth process, cylindrical holes of about \mbox{1 $\mu$m} in diameter are formed and act as scatterers and provide the optical feedback for the random lasing. The scatterer density varies over the sample but can be estimated roughly to a few hundred per square millimetre. From these parameters, a mean free path of  some millimetre is estimated, which is much larger than wavelength and still larger than the length of the active area defined by the excitation spot under optical pumping. Hence, we are studying a weakly scattering configuration far out of the regime of Anderson localization and also far from the diffusive regime \cite{caoreview2005}. While the capability of random lasing of the air-hole/gain configuration has been demonstrated previously \cite{kalus11}, the following time-resolved experiments will give some insight of the temporal behaviour of this random lasing system.

%%%%%%%%%%%%%%%%%%%%%%%%% exp setup

%\subsubsection{Setup}
The experimental setup is presented in figure \ref{fig:setup}. The sample is optically pumped by amplified spontaneous emission of the laser dye 2-Methyl-5-t-butyl-p-quaterphenyl (DMQ)  with a single laser mirror to increase the output of the emission. The dye in turn is pumped by a XeCl excimer laser (lambda physics) at a repetition rate of 1 to \mbox{10 Hz} with a wavelength of 308 nm. This provides a temporally smooth pulse of 10 ns  with a spectral maximum at \mbox{363 nm} and a width (FWHM) of about \mbox{3 nm.} To compensate for the high divergence of the amplified spontaneous emission (ASE), a \mbox{600 mm} lens is placed with its focal plane in the dye cuvette to collimate the beam. The pump pulse is focused on a slit by a combination of a spherical and a cylindrical lens. An image of this slit is projected on the sample by a \mbox{80 mm} biconvex lens to a stripe of about \mbox{2 mm $\cdot$ 0.5 mm.} The emission from one edge of the sample is collimated by and focused with two \mbox{80 mm} lenses on the entrance slit of a small Rowland type monochromator. This enables us to identify different lasing modes in the optical spectra with a resolution \mbox{$ < $0.1 nm.} The temporal evolution of the spectrally resolved emission is investigated by a Hamamatsu streak camera (Model C5680) in combination with the single shot unit (M5676) to make a triggered single sweep possible. The streak camera is triggered by an electronic pulse generated by the control computer of the excimer laser. This pulse is delayed by a pulse generator to achieve a temporal overlap of the emission and the sweep by the streak camera. A sapphire substrate is placed in the pump beam to reflect a reference signal which is directly projected on the streak camera to enable a temporal comparison of excitation and emission. All measurements are done at room temperature. 

%%%%%%%%%%%%%%%%%%%%%%%%% exp timeslice
\begin{figure}[ht]
\centering
\includegraphics[width=8cm]{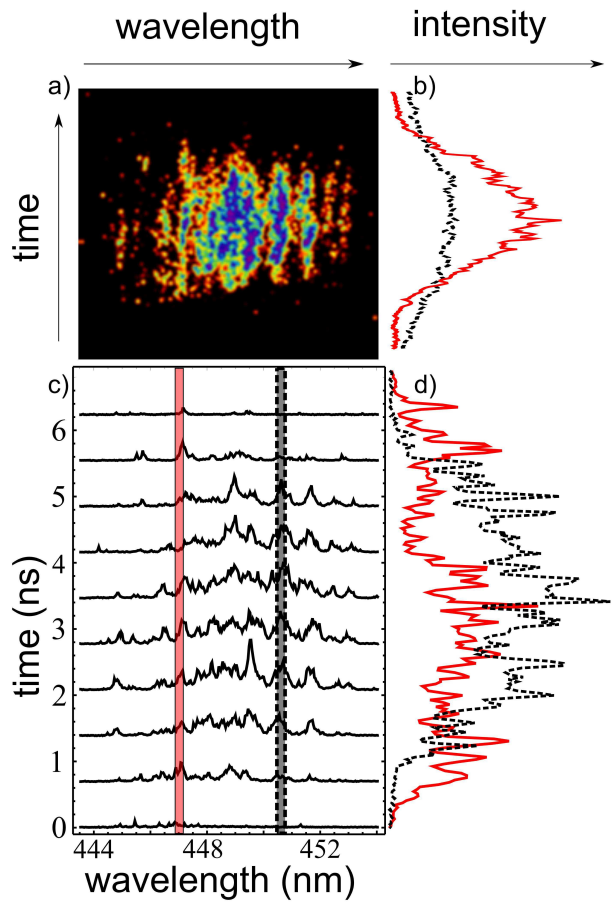}
\caption{ \label{fig:streak1}
a) 
Typical streak camera image of the random laser emission smoothed by a Gaussian filter over 2 pixels to make the spectra less noisy. It is colour coded with black for low over red, yellow, green to the highest intensities in blue.  
b) 
Spectrally integrated time profile of the emission (red) and the excitation pulse (black, dashed). 
c) 
Slices at every 0.7 ns through the spectrum integrated over the respective interval (0 ns to 0.7 ns, 0.7 ns to 1.4 ns,...). 
d)
Temporal profile of the two modes marked in c).
}
\end{figure}

Figure \ref{fig:streak1} characterizes the emission of a representative single shot. The pump energy is about \mbox{10 mJ/cm$^2$}. Panel a) visualizes the primary data averaged with a Gaussian filter over two pixels to reduce noise and make the modes more distinct. The other panels are different representations of these data to emphasize different characteristics of the emission.  Panel b) shows the temporal variation of the spectrally integrated emission (red) and excitation (black, dashed) intensities. Both curves are smooth with a similar rise and fall behaviour. The emission starts slightly later and stops earlier then the pump pulse, when the RL passes threshold. The evolution of spectra shown in panel c) clearly exhibits multiple modes at all instants of time. 
They have randomly distributed spectral positions in qualitative accordance with most other RLs as mentioned in the introduction. The positions of the modes stay constant over all times but their intensities evolve differently. To make the last point more obvious, panel d) compares the intensity evolution of the two modes marked by red and black (dashed) stripes in c). No temporal averaging is done to make also the fluctuations of the primary data visible. The red mode starts earlier and ends later as the other one, what means it has a lower threshold. After about one nanosecond, the black (dashed) mode also passes threshold, increases rapidly and exceeds the red one by a factor of about two. In the middle part of the pulse, the black (dashed) mode stagnates and the red one makes up. This sequence of events inverts in the falling part of the pump pulse. 
Only reference  \cite{Soukoulis2002} reports measurements with comparable simultaneous temporal and spectral resolution.  There, similar multiple modes with individual temporal behaviour have been observed. However, the extremely short 20-ps excitation would cause such multi-mode emission even in conventional Fabry-Perot configurations. In contrast, the 10-ns pump ramp of our experiment can be regarded as quasi-stationary excitation. Our results are the first experimental confirmation of the effects of mode competition which have been theoretically derived for static pump levels \cite{Tureci2008a,tureciab2009,Andreasen2011}.

%%%%%%%%%%%%%%%%%%%%%%%%% exp multiple shots
\begin{figure}[ht]
\centering
\includegraphics[width=9cm]{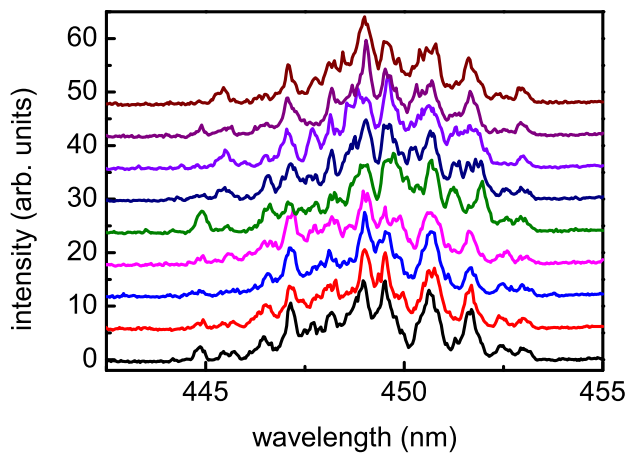}
\caption{ \label{fig:multishot}
Comparison of nine different shots with pump energies of 9 to 10 mJ/cm$^2$. The time integrated spectra show the edge emission under identical focus and detection conditions.
The lines are vertically shifted for better visibility.
}
\end{figure}

So far we have considered the mode dynamics during a single shot.  Next question is whether the mode structure is stable from shot to shot, as can be expected for a fixed scatterer distribution.  Figure \ref{fig:multishot} shows the time integrated spectra of nine different shots with a pump energy of 9 to 10 mJ/cm$^2$.  Within the experimental uncertainties, the mode picture stays stable from shot to shot. However, this happens only if the excitation conditions do not change between the shots. This stability has only been achieved by using highly reproducible ASE excitation pulses. The fluctuations of pump-laser pulses from a standard dye laser were too large for these purposes.  We conclude that the evolution of the modes is completely determined by the scatterer configuration but depends sensitively on the excitation conditions.  Latter fact can perhaps explain the different results of reference   \cite{mujumdar2007}, where the mode spectra vary from shot to shot although unchanged far-field speckle  patterns  of the pump light proof a stable scatterer configuration. Intensity fluctuations of the pump-laser pulses are not excluded by invariant speckles.

% Model
\section{Model} \label{sec.model}
%------------------------

To model the dynamics of the present RL, some difficulties have to be faced. First, the direct numerical solution of semiclassical Maxwell-Bloch type equations \cite{Andreasen2011,sebbahrandom2002,Liu2009,Liu2007} is limited to small sample sizes up to few hundred $ \lambda^2 $.  A different approach is required for the sizes-scale of our experimental configuration, which exceeds $ 10^7 \lambda^2 $. This problem is solved by approximations that exploit the large distances between scatterers. A second difficulty originates in the random nature of the scatterer distribution. The specific configuration depends on the position of the pump spot on the sample and is not known in detail. Therefore, we calculate series of different realizations and demonstrate their common qualitative features by means of characteristic examples.

Basic constituents of the model are $ N $ {scatterers} in the gain carrying planar wave guide with $2N(N-1)$ directed in-plane optical pathways between them, which we call {rays}.  
The optical amplitude of the light travelling along a ray $ {ij} $ (from scatterer $ j $ towards scatterer $ i $) is
\begin{eqnarray}\label{field}
{E}_{ij}(z,t) = \textrm{Re}\left\{{\cal E}_{ij}(z,t) G(kz) e^{-i \omega_0 t }\right\},
\end{eqnarray}
where $ z $ is the distance travelled on the ray. The far-field 2D Green function $ G(kr)=\exp(ikr+i \pi/4)/\sqrt{8 \pi kr}$ describes static amplification and phase shift of a scattered wave in a medium with a spectrally constant complex reference wave number $ k = \bar n \omega_0 / c - i (\bar g-\alpha_0)/2 $ ($\bar n, \bar g, \alpha_0 $: reference values of refractive index, gain, background losses). The dynamics is contained in the slow amplitudes, which we assume to obey
\begin{eqnarray}
\label{propagation}
(\partial_z+\frac{1}{c}\partial_t){\cal E}_{ij}(z,t) &= \left[\frac{1-i \alpha}{2} (g(z,t) - \bar g) -\frac{ \alpha_0 }{2}\right]{\cal E}_{ij}(z,t),
%~~~~~ \textrm{ where } g(z,t)=g_d(t) \textrm{ if} z\in d,
\end{eqnarray}
where $ g(z,t) $ denotes the local gain coefficient.  The term with the $ \alpha $-factor is the standard model for the amplitude-phase coupling in a semiconductor laser.   The plane-wave propagation equation (\ref{propagation}) holds in good approximation because the scatterers are separated by some ten $ \mu $m, which  is much larger than both wavelength (few hundred nm) and scatterer size ($\approx 1  \mu $m).  It is an appreciable simplification because the optical field needs to be calculated only on the network of straight lines between scatterers but not in all area. Scattering enters via the boundary conditions
\begin{eqnarray} \label{b.c.}
{\cal E}_{ij}(0,t)=\sum_{j'} A_{ijj'} \left[
       {\cal E}_{jj'}(l_{jj'},t)G(kl_{jj'})  +  \beta_\textrm{spont} 
      \right].
\end{eqnarray}
The scattering amplitudes $ A_{ijj'} $ from ray $ jj' $ into ray $ ij $  depend in general on the scattering angle, because the scatterer are larger than the wavelength (Mie scattering). $ l_{jj'} $ denotes the length of ray $jj'$. $ \beta_\textrm{spont} $ is a small Langevin force simulating spontaneous emission impinging on the scatterers from everywhere. Other noise sources are disregarded for simplicity. 

Calculating the local gain $ g(z,t) $ requires an equation for the occupation inversion.  The stimulated emission therein has strong sub-wavelength variations due to multi-wave interferences of the optical intensity (cf. calculated intensity distributions in references \cite{sebbahrandom2002,Liu2009}) .  In semiconductors, these variations are smoothed by the diffusion of charge carriers. Taking this effect implicitly into account, we partition the pumped area appropriately into domains $ d $ larger  than the diffusion length, represented by a spatially averaged gain $ g_d(t) $. The dynamics of $g_d(t)$ is modeled by the rate equation
\begin{eqnarray}\label{inversion}
\tau_n \frac{\rmd}{\rmd t} g_d(t) &= g_0(t) - g_d(t) \left[1+S_d(t)\right],
\end{eqnarray}
with inversion life time $\tau_n$ and unsaturated gain $g_0(t)$ (pump 
term). $ S_d(t) $ is the average intensity in domain $ d $.

Equations (\ref{propagation}) to (\ref{inversion}) are the core of our model. They are the dynamical generalization of the steady-state RL model of Ref. \cite{kalus11}, which is re-obtained when assuming constant $ g(z,t) $ and $ \beta_\textrm{spont}=0 $. 

\section{Exemplary Simulations} \label{sec.simulation}
%--------------------------------

\begin{figure}[ht]

\centering
\includegraphics*[width=10cm]{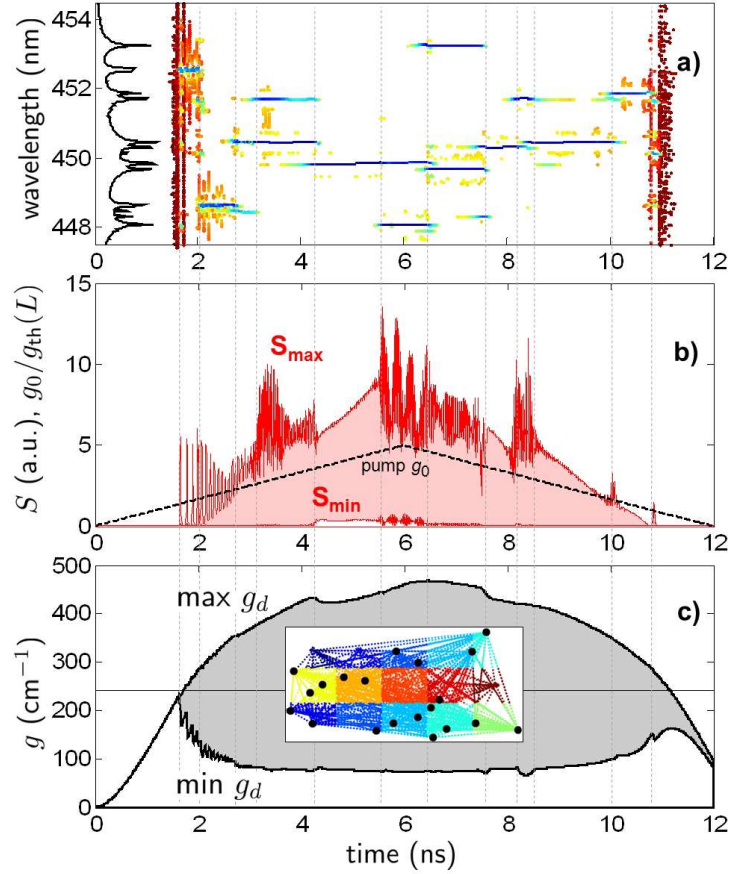}
\caption{\label{fig.simufig}
An exemplary simulation result. 
a):
Optical spectra. Black line: time-integrated spectrum (log scale, arbitrary units, rotated $ 90^o $). Coloured dots: peaks of the optical spectra vs. time (thick blue: high intensity, thin red: low intensity). Spectra are calculated with a shifting window of length $2^{11} dt \approx 77$ ps. For each window, the positions of spectral peaks are plotted at the centre of the window. Logarithms of the spectral peak heights are colour coded from dark blue = highest peak in the actual window to red = 40 dB less. Peaks below -44 dB are disregarded. 
b): 
Variation of pump (unsaturated gain $g_0$ in Equation (\ref{inversion}), relative to threshold, black dashed) as well as maximum and minimum of optical intensity within the pumped region (red). 
The black dashed line and the upper red line correspond to the measured pump (black, dashed) and emission profile (red), respectively, in figure \ref{fig:streak1}b of the experimental section. 
c): 
Transients of maximum and minimum of gain $g_d$ in pumped region. Black horizontal: $g_\textrm{th}(L)$, equation (\ref{pairthreshold}).  
Inset: the considered exemplary configuration of 20 scatterers (thick dots) within a  0.4$\times$0.2 mm$^2$ pumped stripe. Thin coloured dots between the scatterers represent the numeric grid along the rays. Different domains are coded by different colours. 
}
\end{figure}

We have solved the model equations for several different configurations of randomly distributed scatterers.  The numerical schema is briefly described in the appendix. Multi-mode operation qualitatively similar to experiment is obtained in all cases. One particular configuration can thus serve as representative in the following. 20 point-scatterers are randomly positioned in a 0.4 mm times 0.2 mm excitation stripe as sketched in the inset of figure \ref{fig.simufig}c. This ensemble is slightly smaller than the experimental ones but better suited for visualizing its internal structure. 
Further parameters are: 
Central vacuum  wavelength $ \lambda_0= 450$ nm. 
Phase and group velocities $c=c_0/\bar n$ with $ \bar n = 2 $. 
Inversion life time $ \tau_n = 500 $ ps.
Amplitude-phase coupling $ \alpha=-5 $. 
For simplicity, we use the isotropic scattering amplitude $ A = 4i $, the strongest possible elastic point-scattering. This avoids the necessity to discuss dependencies on scatterer size and the resonances related to it, which are of minor interest in the present context. The corresponding scattering cross section is $ \sigma = |A_s|^2\lambda_0/8 \pi \bar n \approx 100$ nm. With the scatterer density $ \rho \approx 250 \textrm{ mm}^{-2}$, the corresponding mean free path of light $ l_\textrm{free} \approx 1/(\sigma \rho)  \approx 4$ cm is much larger than the size of the excitation spot. Thus, this point-scattering configuration belongs to the same class of RL as our experimental realization. Both differ from other RLs, where the scatterers are much closer to each other (see, e.g., \cite{wiersma08,zaitsev10,wurandom06}). The threshold of a weakly scattering RL is governed by the two scatterer with largest separation $L$  \cite{wurandom06,kalus11}, namely 
\begin{eqnarray} \label{pairthreshold}
g_\textrm{th}(L) = \frac{1}{L}\ln (\frac{1}{R_\textrm{eff}})~~~~ \textrm{ with effective reflectivity }
R_\textrm{eff} = \frac{\sigma}{ 2 \pi L}.
\end{eqnarray}
This threshold is as high as about 250 cm$^{-1}$ due to the extremely small feedback $R_\textrm{eff}\approx 4.6 \times 10^{-5}$. 

%%%%%%%%%%%%%%%%%%%%%%%%% simufig

The pump pulse used in the numerical calculations increases within 12 ns from zero to five times threshold and back. The simulation results are summarized in figure \ref{fig.simufig}. The time-integrated spectrum at left vertical axis in panel a) exhibits the irregular multi-mode structure typical for RLs. Different modes stem from different time intervals (right part of figure), in qualitative agreement with our experimental results. The modal structures during the pump-down part of the excitation pulse are not completely symmetric to the pump-up part. This feature indicates possible multi-stabilities or rather long time scales of mode competition. Depending on which modes are active, the transient can be divided into different epochs (indicated by thin vertical dashed lines). First, the gain rises staying spatially homogeneous until exceeding threshold. Lasing of a mode at $\lambda\approx 452.6$ nm starts with a series of damped relaxation oscillations (RO). Similar RO have been obtained in all calculated configurations but not observed in experiment. This discrepancy may be due to underestimation of damping of the RO by neglecting nonlinear gain saturation in the model and the limited temporal resolution in the experiment.  In the lasing regime, the stimulated emission makes the gain inhomogeneous (panel c). The gain is strongly depleted mainly at the ends of the laser, where the intensity is large, whereas it continues growing in the middle of the pumped area. The ratio between largest and smallest gain reaches values as high as five. Reason for this strong spatial hole burning (SHB) is the extreme amplification along the stripe, which es required for overcoming the large scattering losses. Large SHB is well known from Fabry-Perot lasers with small reflectivities. In the present case, increasing SHB reduces the gain of the lasing mode until another mode at $ \lambda \approx 448.7 $ nm takes over few hundred picoseconds after onset of lasing. The  hole burning deepens further, leading to a series of further mode jumps. Most mode jumps are accompanied by comparatively sudden changes of gain and power. The short-time variation of the power in certain epochs is due to the fast beating of two or more active modes. These variations disappear in epochs with single-mode operation.   They are not resolved in experiment due to limited band width.

\section{Network Aspects} \label{sec.network}
%------------------------

Let us regard now the model above in terms of a network with the scatterers as nodes and the rays as links. This network is fully connected and static because each scatterer is always linked to every other one. The dynamics is carried by the amplified streams of light along the links and their redistribution by scattering at the nodes. In what follows we evaluate this dynamics by introducing an appropriate time-dependent weight $ w_{ij}(t) $ to each link from $j$ to $i$, which measures its importance for the network. Weighted networks have been successfully used to analyse other transport scenarios, e.g., road traffic \cite{bonoroad2010} or international trade streams \cite{fagiolo2009}. Of course, the results will depend on how the weights are chosen. Here, two different weights will be compared, based on the amplification along a ray and the intensity of the light stream on it, respectively. 

\subsection{Weighting links by amplification}
%----------------------------------------------------

\begin{figure}[ht]

\centering
\includegraphics*[width=10cm]{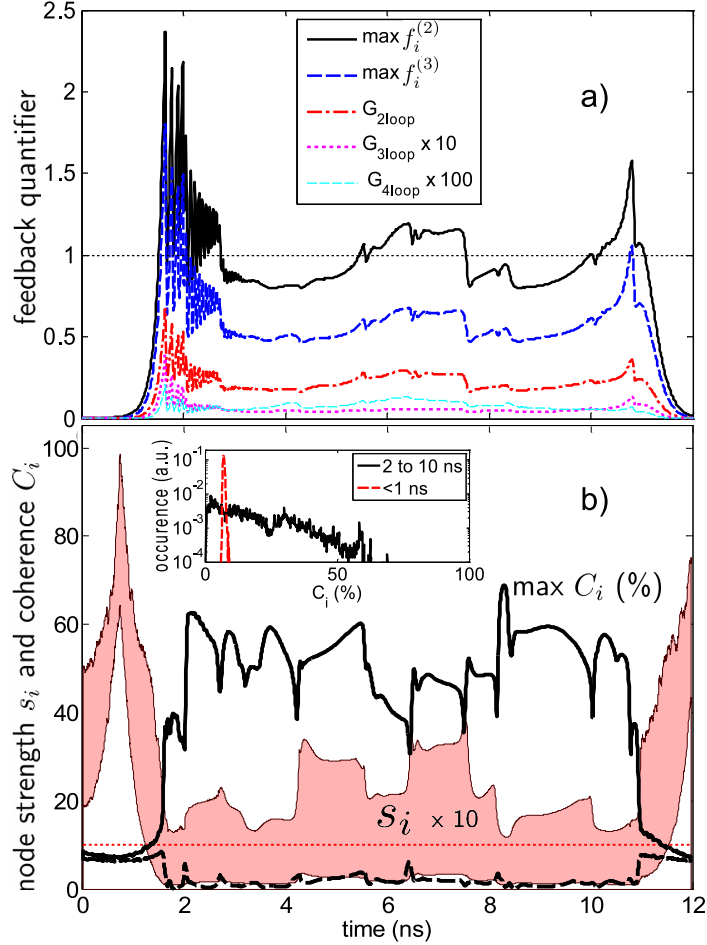}
\caption{ \label{fig.gnetfig}
Variation of weighted-network quantifiers in the time interval of figure \ref{fig.simufig}. 
a)
Using net amplification (\ref{weights}) as weights. 
Red dash-dotted, magenta dotted, and cyan thin dashed: 
maximum amplification along lowest-order n-loops.
Solid black and blue dashed:
maximum summarized direct and 3-loop feedbacks of a scatterer according to Equations (\ref{fn}).
b)
Using intensity as weight, Equation (\ref{wije}).
Red-grey shaded: 
range between minimum and maximum of node strength $s_i$, Equation (\ref{sie}), multiplied by ten. The horizontal red-dotted line indicates the level $ s_i=1 $. 
Black solid and dashed: 
maximum and minimum of scatterer coherence $C_i$, respectively, in \%, Equation (\ref{cie}). The inset shows the distributions of coherence below threshold (red dashed) and above threshold (solid black).
}
\end{figure}

In our lasing network, the magnitude of amplification along a link,
\begin{eqnarray} \label{weights}
w_{ij}(t) = \left|A \cdot G(kl_{ij}) \exp\left(\frac{1}{2}\int_{ij} g(z,t) ~ dz\right)\right|, 
\end{eqnarray}
is a natural choice of its weight. A single scattering event is included here by the scattering amplitude $ A $. This weight becomes unity if the amplification compensates the scattering losses. For weak scattering,  the laser condition is well approximated by the pair-threshold (\ref{pairthreshold}) \cite{kalus11}, which corresponds to $\max w_{ij} =1$. Thus, this choice of weights prefers those links which are most important for the lasing. Dynamics is brought  in the otherwise static network of fully connected scatterers by the evolution of weights. It is hopeless to consider them individually.  Appropriate summary quantifiers are required. Numerous such quantifiers have been used in literature, see e.g. \cite{RevModPhys.74.47,onnela05,boccaletti2006,dorogovtsev08}. Inspired by these ideas, we consider the following choice adapted to laser physics:
\begin{eqnarray} \label{fn}
f^{(n)}_i= {\sum_{i_2,\ldots,i_n}}'w_{ii_{2}}w_{i_2i_3}\cdots w_{i_ni}, 
~~~~(n=2,3,\ldots).
\end{eqnarray}
These quantities represent the summed magnitudes of feedback from all $ n $-loops beginning and ending at scatterer $ i $. An $ n $-loop is a closed path through $ n $ nodes. It is irreducible, i.e., each node is touched only once, which is symbolized in (\ref{fn}) by the prime at the sum. Note that this summation of magnitudes describes a fictive totally constructive superposition of light returning back from all different $ n $-loops. It is the maximum possible $n$-loop feedback for the given gain distribution. $ f^{(2)}_i $ and $ f^{(3)}_i $ correspond to the node strength and the cluster coefficient, respectively, often used in the analysis of weighted networks \cite{boccaletti2006}. Their largest values among the different scatterers $ i $ as calculated from the simulation results are plotted against time in figure \ref{fig.gnetfig}a. The largest $  f_i^{(2)} $ reaches unity at threshold, oscillates during the RO and remains $\approx 1$ in all epochs with pump above threshold. This supports the idea that the light circulating on the 2-loop between the most distant scatterers plays a dominant role. However, even after the RO, the deviations from unity are not negligible. The pump changes quasi statically here and the laser operates close to threshold condition. A maximum $ f_i^{(2)}  $ $<1$ indicates that larger loops must also contribute to the feedback. Indeed, the maximum 3-loop feedback is already sufficient to fill the gap. This means, the total feedback is composed of superpositions of several different loops. In order to get further insight into the role of closed light-loops for the operation of the RL we display in panel a) also  the  maximum amplifications along single irreducible $ n $-loops,
\begin{eqnarray} \label{Gloop}
G_{n\textrm{\footnotesize loop}}= {\max}'w_{i_1i_{2}}w_{i_2i_3}\cdots w_{i_ni_1}.
\end{eqnarray}
The prime at the max again symbolizes that no node index occurs twice. $G_{2\textrm{\footnotesize loop}}$ is about three to four times smaller than unity, i.e., coherent superposition of at least three different 2-loops is acting in these epochs. The maximum feedback along higher $n$-loops is very small and decreases rapidly with $n$. We can conclude that a single closed loop of light can not carry the lasing completely in our case. However, the decrease of $G_{n\textrm{\footnotesize loop}}$ with $n$ can become overcompensated by the rapid growth of  the number of $n$-loops. Indeed, calculated $ f^{(n)}_i $ increase with $n$ for $n>2$. It is questionable, whether this behaviour reflects the real role of $ n $-loop feedback, because it can be assumed that the phases of light returning from different loops differ from each other, causing a high degree of cancellation, which is disregarded in the summation of magnitudes in $ f^{(n)}_i $. Obviously, the phase-insensitive weight (\ref{weights}) is no good measure for multiple feedback effects.

%%%%%%%%%%%%%%%%%%%%%%%%% quantifier

\subsection{Weighting links by optical intensity}
%--------------------------------------------------------
Now we regard the rays between scatterers as links that are transporting optical fields and choose the weight proportional to the optical intensity on the link, 
\begin{eqnarray} \label{wije}
w_{ij}(t) = \frac{S_{ij}(t)}{\max S_{ij}(t)}
~~~ \textrm{with}~~~
S_{ij}(t)=\langle |{\cal E}_{ij}(t)G(kl_{ij})|^2\rangle, 
\end{eqnarray}
where for shortness $ {\cal E}_{ij}(t) $ denotes the amplitude at the end of the ray. The normalization makes the strongest link having weight 1. The angle bracket $ \langle \cdots \rangle $ denotes averaging over a 50 ps time interval in order to suppress possible fast oscillations due to mode beating. Being the intensity at the end of the link, this weight includes the amplification along the link as well. But it also depends on the amplitude at the beginning, given by the superposition (\ref{b.c.}) of complex scattered amplitudes. This way, it contains information on optical phases. The strength of a scatterer in the network is measured by the sum of all impinging intensities as 
\begin{eqnarray}\label{sie}
s_i= \sum_{j\ne i} w_{ij}. 
\end{eqnarray}
It reflects the effective number of links, which the node is connected to in the network. The maximum $ s_i=N-1 $ is reached only if all links have the same weight. 
High impinging intensities can however get useless in case of destructive interference. In order to have an explicit measure also of this phase-sensitive process, the additional quantifier
\begin{eqnarray} \label{cie}
C_i(t)= \left\langle ~ \left(\frac{\left|\sum_j{\cal E}_{ij}(t)G(kl_{ij})\right|}
                          {\sum_j\left|{\cal E}_{ij}(t)G(kl_{ij})\right|}
        \right)^2~\right\rangle 
\end{eqnarray}
is introduced that we call the coherence of the scattering at $i$.
It ranges between $0$ and $1$, depending on whether the impinging fields superpose destructively or constructively, respectively. 

Panel b) of Figure \ref{fig.gnetfig} illustrates the evolution of these quantifiers. The different epochs of figure \ref{fig.simufig} are clearly resolved. In particular, a striking qualitative difference appears between sub-threshold and lasing regimes. Before the onset of lasing, all scatterers have nearly the same coherence of about 6\%. This is close to the corresponding coherence $ 100\%/(N-1) = 5.3\%$ of $ N-1=19 $ impinging rays with equal magnitude and random phase and, thus, the fingerprint of the dominating spontaneous emission. In the same interval of time, the node strengths $s_i$ exhibit an unexpected maximum at about half the turn-on time. Here, the effective number of links of the nodes gets maximum, ranging from 6 to 10 of $ 19 $ possibilities. The network is most connected here. The physics behind is the competition between the geometrically determined decrease of the amplitude of a circular scattered wave and its increase due to optical amplification. In the initial moments of time, amplification is negligible and the shortest links have the largest weight. Accordingly, $s_i$ is roughly the effective number of next neighbours. With progressing time, amplification increases, and the amplitudes impinging from farthest nodes grow. This means an effective equalization of the weights and an increase of $s_i$. Beyond a certain level of amplification, the long-distance amplitudes dominate, the equalization diminishes, and $s_i$ falls again. This effect is typical for a RL, at least in case of weak scatterers. 

In the lasing regimes, the situation changes dramatically. The largest $s_i$ varies between closely above one and about four. This means, large intensities are carried by only few links. The lowest $s_i$ are nearly zero, i.e., there are scatterers which remain nearly unilluminated. The coherence behaves similarly. Its maximum is drastically enhanced compared to the regime of spontaneous emission. Values between 40\% and 60\% allow for a highly  efficient superposition of the impinging waves. The minimum on the other hand becomes very small. At those nodes, the impinging waves cancel each other by destructive interference. They are sinks of radiation. The histogram of $ C_i(t) $ in the inset of panel b reveals another surprise: when lasing starts, a considerable part of the nodes ad as sinks. This feature points to an interesting way of self-organization in the RL: few scatterers are selected by a high degree of coherence, whereas many others become devitalized by destructive interference.

\begin{figure}[ht]
\includegraphics*[width=\textwidth]{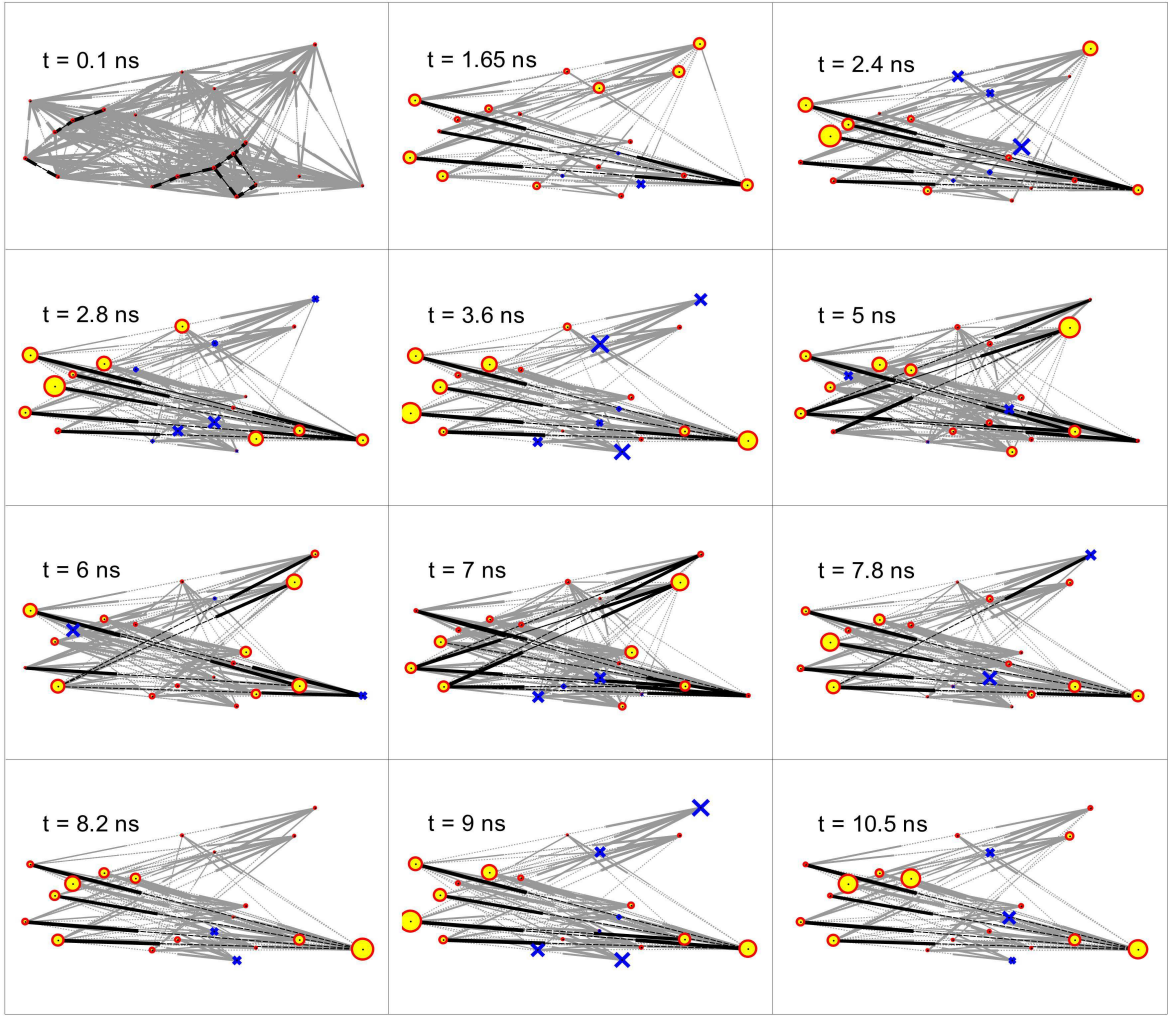}
\caption{ \label{fig.modennetz}
Top views on the RL-network. The instants of time $t$ are representative for the different epochs sketched in figure \ref{fig.simufig}.
Dots: 
positions of scatterers. The sizes of coloured (grey) circles around the scatterers are proportional to their coherence $C_i(t)$. 
Blue crosses:
scatterers being radiation sinks, i.e. $ C_i(t)<0.05 $. The size of the crosses is proportional to  $( 0.05-C_i) $, i.e., the largest crosses indicate $ C_i\approx 0 $.
Lines:
links between scatterers. The lines are split in three parts. The thickness of the part close to a scatterer is proportional to the logarithm of the intensity impinging on the scatterer, i.e. of the weight $w_{ij}$ according to Formula (\ref{wije}). Weights above 50\% are black, all others grey. Weights below 1\% are disregarded. 
}
\end{figure}

In order to get more insight into this process, figure \ref{fig.modennetz} presents spatially resolved graphical representations of weights and coherence for moments of time representing the different epochs of figure \ref{fig.simufig}. Obviously, the distribution of intensity \mbox{( = weight)} over the links and the distribution of coherence among the scatterers differ between the epochs. Only the panels at 3.6 ns and 9 ns agree nearly completely with each other. They belong to the same dominant mode at \mbox{$ \lambda= $ 450.5 nm} (cf. figure \ref{fig.simufig}a). The small differences might be due to different intensities of the side mode at \mbox{451.7 nm}.  The most obvious differences appear between subthreshold-panel  \mbox{$ t=0.1 $ ns} and all other ones, which belong to lasing states.  Thus, turn-on and switch-off of the laser appear as the most drastic reorganizations of the underlying network. Below threshold, the coherence of all nodes is small, amplification is negligible and the shortest links carry largest intensity (black lines), whereas the longest ones do so in the lasing cases.

%%%%%%%%%%%%%%%%%
%%%%%%%%%%%%%

Although differing from each other in detail, the panels belonging to the lasing state also exhibit some similarities. The strongest links (black) always connect left scatterers with right ones and their length is comparable to the longest pair distance. There is no indication of strong 3-loops or even higher loops, as expected in this weak-scattering regime. The number of strong links is always larger than unity but small compared to the total number of links. Strong links are always connected with each other, i.e., every scatterer on a strong link can be reached along strong links from any other scatterer on a strong link. This illustrates that those modes are lasing, which are able to establish several coupled links with strong amplification coherent to each other. This can also be seen from the depicted coherences: scatterers with high coherence belong mostly to strong links. However, there are also interesting exceptions. At \mbox{7.8 ns}, e.g., the top right scatterer receives a high intensity from one on the left bottom, but its coherence is close to zero. As a consequence, the amplitude scattered back to the left bottom node is tiny, the scatterer is ignored by the network. Similar destructive sinks of radiation appear also in other epochs (blue crosses).  The threshold is smallest here for modes at wavelengths that exclude feedback from a part of scatterers.  It is apparently impossible to incorporate these scatterers coherently in the network. They are somehow frustrated similar to the frustration of certain particle packings in condensed matter \cite{Nelson1983}. 

\section{Summary and conclusion} \label{sec.conclusion}
%------------------------------------------

Combining time-resolved spectroscopy, numerical simulation and network analysis,  a deepened picture of the mode dynamics in a semiconductor RL with  dilute weak scatterers has been obtained.  In contrast to other experimental work, which uses pulsed laser excitation,  the amplified spontaneous emission of a laser dye provides the necessary smooth and reproducible pump pulses to investigate the RL in an quasi-stationary state. The time-resolved optical spectra exhibit multimode spectra with individually developing modes but no variation from shot to shot. These findings are qualitatively reproduced with a numerical model. Exploiting the specifics of the dilute system of weak scatterers, this model is mapped to a lasing network. To quantify the dynamics of the network, the node coherence is introduced as a new quantifier that takes into account the phase of the laser light.  Its use has revealed new aspects of the self-organization of the laser field. Lasing is carried by connected links between a subset of scatterers, the fields on which are oscillating coherently in phase. In addition, perturbing feedback with possibly unfitting phases from frustrated other scatterers is suppressed by destructive superposition. We believe that our findings are representative at least for weakly scattering RLs. The generalization to RLs with more dense, stronger and also active scatterers should be possible when basing the model on a more complex scattering theory \cite{Felbacq1994} or the recent Euclidean matrix theory \cite{Goetschy2011}.

\ack

This paper was developed within the scope of the IRTG 1740 /
TRP 2011/50151-0, funded by the DFG / FAPESP. 
The authors also like to thank Sergey Sadofev for the sample growth.

\appendix

\section*{Appendix: Numerical Implementation} \label{sec.numerics}
%------------------------------------
\setcounter{section}{1}

To calculate spectral dynamics within a given spectral interval $ \Delta \lambda $ (7 nm), centred at a wavelength $\lambda_0$ (450 nm), a time step $ \rmd t = \frac{ \lambda_0^2 }{2c_0 \Delta \lambda} $ (48 fs) is used. It yields the space step $ \rmd l = c \rmd t $ (7.2 $\mu$m) for discretizing on each ray.  The last grid point of the ray is taken just beyond or at the final scatterer.  On the grid, the field   is represented by logarithmic amplitudes
\begin{eqnarray} \label{psi}
\psi = \ln({\cal E}(z,t))
\end{eqnarray}
for numerical efficiency. The $ \psi  $ of all rays are put subsequently into a large linear array. 

In each time step, $ \psi $ and $ g_d $ are updated in sequence. The update of $ \psi $ is done in two subsequent steps. First, the field is propagated according to (\ref{propagation}) one step along the rays
\begin{eqnarray} \label{psiupdate} \fl
\psi(z,t)=\psi(z-\rmd l,t-\rmd t) + \frac{1}{2}\left[(1-i \alpha)(g_{d}(t-\rmd t/2)-\bar g)-\alpha_0\right] \rmd l.
\end{eqnarray}
This propagation step is most time consuming. Best performance is found when exploiting the matlab operation {\sc CircShift}, which shifts the components of an array circularly. New in-values on each ray are set in a second step according to 
\begin{eqnarray} \label{Ein} \fl
\psi_{ij}^\textrm{in} = \ln( {\cal E}_{ij}^\textrm{in}), ~~~\textrm{  where } ~~~
{\cal E}_{ij}^\textrm{in}(t)=\sum_{r'} A_{r r'} \left[
       {\cal E}_{r'}^{out}(t)G(k,l')  +  \beta_\textrm{spont} 
      \right]
\end{eqnarray}
are the fields injected into ray $ r $ by the instantaneous scattering.  Since in general the final scatterer of a ray is placed between the two last grid points, this amplitude is not exactly available but is determined by linear interpolation. This approximation introduces an artificial numeric dispersion that suppresses high-frequency modes. We did not find a way to better treat this problem. Fortunately, this numeric dispersion can be used to simulate the real gain dispersion, which is limiting the amplification band width. 

Now, the gain $ g $ is updated. It is given on the independent spatial domain-grid. The temporal grid for $g_d$ is shifted by $\rmd t/2$ compared to the grid of $ {\cal E} $. Accordingly, the simplest integration of the rate equations (\ref{inversion}) over one interval $\rmd t $ yields the update formula
\begin{eqnarray}
    g_d(t+\frac{\rmd t}{2})=g_d(t-\frac{\rmd t}{2})+\left[g_0-g_d(t-\frac{\rmd t}{2})(1+S_d(t))\right] \frac{\rmd t}{ \tau_n }.
\end{eqnarray}
The mean intensity $S_d$ in the domain is estimated as the average over the impinging intensities at all scatterers in the domain, symbolically
\begin{eqnarray}
S_d = \langle |E_s|^2 \rangle_{s \in d}.
\end{eqnarray}
Note, the $E_s$ contain the Green function $G$ in contrast to the prefactor $ {\cal E} $. 

The described numerical approach is programmed with Matlab. The presented example required about 1.5 minutes runtime for \mbox{1 ns} simulated time on a Dell PowerEdge T710.


\section*{References}


\begin{thebibliography}{10}

\bibitem{caoreview2005}
Cao H 2005
\newblock {Review on latest developments in random lasers with coherent feedback}
\newblock {\em \JPA} {\bf 38} 10497--535

\bibitem{wiersma08}
Wiersma D S 2008
\newblock {The physics and applications of random lasers}
\newblock {\em Nature Physics} {\bf 4} 359--67

\bibitem{zaitsev10}
Zaitsev O and Deych L 2010
\newblock {Recent developments in the theory of multimode random lasers}
\newblock {\em Journal of Optics} {\bf 12} 024001

\bibitem{polsonrandom2004}
Polson R C and Vardeny Z V 2004
\newblock {Random lasing in human tissues}
\newblock {\em Applied Physics Letters} {\bf 85} 1289--91

\bibitem{knitterspectro-polarimetric2013}
Knitter S, Kues M and Fallnich C 2013
\newblock {Spectro-polarimetric signature of a random laser}
\newblock {\em \PR} A {\bf 88} 013839

\bibitem{follitime-resolved2013}
Folli V, Ghofraniha N, Puglisi A, Leuzzi L and Conti C 2013
\newblock {Time-resolved dynamics of granular matter by random laser emission}
\newblock {\em Scientific reports} {\bf 3} 2251

\bibitem{reddingspeckle-free2012}
Redding B, Choma M A and Cao H 2012
\newblock {Speckle-free laser imaging using random laser illumination}
\newblock {\em Nature Photonics} {\bf 6} 355--59

\bibitem{Heiligenthal2013}
Heiligenthal S, J\"{u}ngling T, D\'Huys O, Arroyo-Almanza D A,
  Soriano M C, Fischer I, Kanter I and Kinzel W 2013
\newblock {Strong and weak chaos in networks of semiconductor lasers with
  time-delayed couplings}
\newblock {\em \PR E}, {\bf 88} 012902

\bibitem{nixon2013}
Nixon N, Ronen E, Friesem A A and Davidson N 2013
\newblock {Observing geometric frustration with thousands of coupled lasers}
\newblock {\em \PRL} {\bf 110} 184102

\bibitem{caorandom1999}
Cao H, Zhao Y, Ho S, Seelig E, Wang Q and Chang R 1999
\newblock {Random laser action in semiconductor powder}
\newblock {\em \PRL} {\bf 82} 2278--81

\bibitem{mujumdar2007}
Mujumdar S, T\"{u}rck V, Torre R and Wiersma D 2007
\newblock {Chaotic behavior of a random laser with static disorder}
\newblock {\em \PR} A {\bf 76} 033807

\bibitem{kalus11}
Kalusniak S, W\"unsche H J and Henneberger F 2011
\newblock {Random semiconductor lasers: scattered versus Fabry-Perot feedback}
\newblock {\em \PRL} {\bf 106} 013901

\bibitem{Baudouin2013}
Baudouin Q, Mercadier N, Guarrera V, Guerin W and Kaiser R 2013
\newblock {A cold-atom random laser}
\newblock {\em Nature Physics} {\bf 9} 357--60

\bibitem{Tureci2008a}
T\"{u}reci H E, Ge L, Rotter S and Stone A D 2008
\newblock {Strong interactions in multimode random lasers}
\newblock {\em Science} {\bf 320} 643--6

\bibitem{tureciab2009}
T\"{u}reci  H E, Stone A D, Ge L, Rotter S and Tandy R J 2009
\newblock {Ab initio self-consistent laser theory and random lasers}
\newblock {\em \NL} {\bf 22} C1--C18

\bibitem{Soukoulis2002}
Soukoulis C, Jiang X, Xu J and Cao H 2002
\newblock {Dynamic response and relaxation oscillations in random lasers}
\newblock {\em \PR} B {\bf 65} 041103

\bibitem{Andreasen2011}
Andreasen J, Sebbah P and Vanneste C 2011
\newblock {Nonlinear effects in random lasers}
\newblock {\em \JOSA} B {\bf 28} 2947--55

\bibitem{sebbahrandom2002}
Sebbah P and Vanneste C 2002
\newblock {Random laser in the localized regime}
\newblock {\em \PR} B {\bf 66} 144202

\bibitem{Liu2009}
Liu H, Liu J, L\"{u} J and Wang K 2009
\newblock {Spectral time evolution of polarized modes under local pumping in a
  two-dimensional random medium}
\newblock {\em \JOA} {\bf 11} 065202

\bibitem{Liu2007}
Liu J S, Xiong Z and Chun W 2007
\newblock {Theoretical investigation on polarization-dependent laser action in
  two-dimensional random media}.
\newblock {\em \JOA}, {\bf 9} 658--63

\bibitem{wurandom06}
Wu X, Fang W, Yamilov A, Chabanov A, Asatryan A, Botten L and Cao H 2006
\newblock {Random lasing in weakly scattering systems}
\newblock {\em \PR} A {\bf 74} 053812

\bibitem{bonoroad2010}
Bono F, Guti\'{e}rrez E and Poljansek K 2010
\newblock {Road traffic: a case study of flow and path-dependency in weighted
  directed networks}
\newblock {\em Physica A: Statistical Mechanics and its Applications}
  {\bf 389} 5287--97

\bibitem{fagiolo2009}
Fagiolo G, Reyes J and Schiavo S 2009
\newblock {The evolution of the world trade web: a weighted-network analysis}
\newblock {\em Journal of Evolutionary Economics} {\bf 20} 479--514

\bibitem{RevModPhys.74.47}
Albert R and Barab\'{a}si A L 2002
\newblock {Statistical mechanics of complex networks}
\newblock {\em \RMP} {\bf 74} 47--97

\bibitem{onnela05}
Onnela J P, Saram\"{a}ki J, Kert\'{e}sz J and Kaski K 2005
\newblock {Intensity and coherence of motifs in weighted complex networks}
\newblock {\em \PR} E {\bf 71} 065103

\bibitem{boccaletti2006}
Boccaletti S, Latora V, Moreno Y, Chavez M and Hwang D 2006
\newblock {Complex networks: structure and dynamics}
\newblock {\em Physics Reports} {\bf 424} 175--308

\bibitem{dorogovtsev08}
Dorogovtsev S N, Goltsev A V and Mendes J F F 2008
\newblock {Critical phenomena in complex networks}
\newblock {\em \RMP} {\bf 80} 1275--1335

\bibitem{Nelson1983}
Nelson D 1983
\newblock {Order, frustration, and defects in liquids and glasses}
\newblock {\em \PR} B {\bf 28} 5515--35

\bibitem{Felbacq1994}
Felbacq D, Tayeb G and Maystre D 1994
\newblock {Scattering by a random set of parallel cylinders}
\newblock {\em \JOSA} {\bf 11} 2526--38

\bibitem{Goetschy2011}
Goetschy A and Skipetrov S E 2011
\newblock {Euclidean matrix theory of random lasing in a cloud of cold atoms}
\newblock {\em Europhysics Letters} {\bf 96} 34005

\end{thebibliography}
\end{document}